\documentclass[draft]{agujournal2019}
\usepackage{url} %
\usepackage{lineno}
\usepackage{soul}

\draftfalse

\journalname{Geophysical Research Letters}

\begin{document}

\title{Deep learning detects uncataloged low-frequency earthquakes across regions}

\authors{Jannes Münchmeyer\affil{1}, Sophie Giffard-Roisin\affil{1}, Marielle Malfante\affil{2}, William Frank\affil{3}, Piero Poli\affil{4}, David Marsan\affil{1}, Anne Socquet\affil{1}}

\affiliation{1}{Univ. Grenoble Alpes, Univ. Savoie Mont Blanc, CNRS, IRD, Univ. Gustave Eiffel, ISTerre, Grenoble, France}
\affiliation{2}{Univ. Grenoble Alpes, CEA, List, Grenoble, France}
\affiliation{3}{Department of Earth, Atmospheric and Planetary Sciences, Massachusetts Institute of Technology, Cambridge, MA, USA}
\affiliation{4}{Dipartimento di Geoscienze, Università di Padova, Padova, Italy}

\correspondingauthor{Jannes Münchmeyer}{munchmej@univ-grenoble-alpes.fr}

\begin{keypoints}
 \item We develop a deep learning-based low-frequency earthquake detection workflow that can detect and locate individual LFEs across regions.
 \item Our approach identifies both previously known and uncataloged LFE sources, showing the higher flexibility compared to template matching.
 \item Deep learning detects LFEs in regions it was not trained on, enabling the search for LFEs in regions without known LFE activity.
\end{keypoints}

\begin{abstract}
Documenting the interplay between slow deformation and seismic ruptures is essential to understand the physics of earthquakes nucleation.
However, slow deformation is often difficult to detect and characterize.  
The most pervasive seismic markers of slow slip are low-frequency earthquakes (LFEs) that allow resolving deformations at minute-scale.
Detecting LFEs is hard, due to their emergent onsets and low signal-to-noise ratios, usually requiring region-specific template matching approaches.
These approaches suffer from low flexibility and might miss LFEs as they are constrained to sources identified a priori.
Here, we develop a deep learning-based workflow for LFE detection, modeled after classical earthquake detection with phase picking, phase association, and location.
Across three regions with known LFE activity, we detect LFEs from both previously cataloged sources and newly identified sources.
Furthermore, the approach is transferable across regions, enabling systematic studies of LFEs in regions without known LFE activity.
\end{abstract}

\section*{Plain Language Summary}
Earthquakes are caused by sudden movements on tectonic faults.
While such sudden movements have been documented for thousands of years, the last decades have revealed that tectonic faults also host a wide range of slow deformation.
Such slow slip happens over the scale of days to years but is still substantially faster than regular plate convergence rates.
Recent years have shown that slow slip can play an essential role in the buildup of large earthquakes.
Classically, slow deformation is detected and characterised using geodetic observations, such as GNSS or InSAR.
This limits the time and space resolution.
An alternative is looking for seismic markers accompanying slow slip, among which the most pervasive are low-frequency earthquakes (LFE).
Due to their low signal to noise ratio and emergent onsets, such LFEs are notoriously difficult to detect.
Here, we develop a novel method for detecting LFEs using deep learning.
Our method successfully detects LFEs from both known and unknown sources.
In contrast to previous approaches, our method can detect LFEs without prior knowledge of the region, which makes it promising for LFE detection in regions where no LFEs have been found previously.

\section{Introduction}

Stress release on tectonic faults can happen in two ways: fast and slow.
Fast deformation happens in the form of earthquakes; slow relaxation is observed as creep or episodes of accelerated slip, so-called slow slip events \cite{dragert2001silent,ozawa2002detection,lowry2001transient,ideScalingLawSlow2007}.
The complex interactions between fast and slow deformation might be at play during the initiation of large earthquakes \cite{radiguet2016triggering,socquetMonthSlowSlip2017a,cruz2021short}.
However, studying these interactions requires detailed catalogs of both deformation types.
While the impulsive nature of earthquakes causes clear signatures on seismic recordings, detecting slow slip is substantially more challenging.
Its detection commonly uses geodetic observations with a limited spatial and temporal resolution \cite{michel2019interseismic,okada2022development,costantino2023slow}.

An alternative way to map slow deformation is by detecting and characterising its seismic markers.
One such type of markers are low-frequency earthquakes (LFEs), weak seismic signals with a duration on the scale of seconds.
Recent research shows that the rate and magnitude of LFEs track the slow deformation \cite{frank2019daily,mouchon2023subdaily}.
LFEs are similar to regular earthquakes in some characteristics, e.g., distinct phase arrivals or predominantly double-couple sources, but also have clear differences \cite{shellyNonvolcanicTremorLowfrequency2007,ide2007mechanism,royer2014comparative,imanishi2016determination,supinoSelfsimilarityLowfrequencyEarthquakes2020,wang2023makes}.
First, they have an eponymous depletion of energy in the high-frequency band (above a few Hz).
Second, in consequence of missing high frequencies, they do not exhibit impulsive arrivals but are emergent, making them hard to detect.
Third, they often occur in intense bursts with inter-event times of only seconds, leading to superimposed waveforms commonly referred to as tremors \cite{shellyNonvolcanicTremorLowfrequency2007}.

To illustrate the challenges these characteristics cause for LFE detection, it is worth contrasting LFE detection with the identification of regular earthquakes.
Detecting regular earthquakes traditionally relies on a two-step procedure: (i) phase picking, i.e., identifying P and S waves arrival times at seismic stations; (ii) phase association, i.e., selecting sets of picks across stations that are consistent with a common source location and origin time.
Downstream analysis can then determine the event location and additional source parameters.
In this workflow, a side benefit of the phase association step is that it acts as a quality control to remove spurious phase picks.
At the moment, such a workflow is not applicable to LFE detection, as the low signal-to-noise (SNR) ratio and the emergent onsets make it impossible for classical algorithms to pick phase arrivals.
Instead, LFE detection usually relies on manual identification \cite{shelly2010migrating}, beamforming \cite{frankAutomaticDetectionLowfrequency2014}, or phase coherence \cite{gombert2023rapid}.
These approaches often suffer from high computational demand or requirement for manual labor.
However, they can be used to generate LFE template waveforms forming an initial catalog for a subsequent matched filtering search on long running recordings.
As the initial approaches often fail to identify all existing LFE sources, such catalogs will be biased towards certain sources.

Due to its high sensitivity, matched filtering, also known as template matching, has become the de facto standard for LFE detection \cite{shelly15YearCatalog2017,bostockMagnitudesMomentdurationScaling2015,frankUsingSystematicallyCharacterized2014}.
Once initial templates are identified, the method identifies repeat occurrences of the template events by correlating these with the continuous waveforms.
In addition to detecting occurrences, this procedure groups the LFEs into families according to their matching templates.
This allows to stack waveforms and accurately locate the families.
While highly sensitive, matched-filtering presents several disadvantages: templates are always region and station specific, matched filtering does not provide locations for individual events, and the model can not detect LFEs outside the initially detected families.
Especially the last limitation shapes our understanding of LFEs, as template matching can only recover repeating events, potentially skewing our view of overall LFE behavior by the most repetitive sources.
Furthermore, the grouping into families is partially artificial, as template matches often overlap, i.e., many detections can not be uniquely assigned to one family.

To address these disadvantages, it would be appealing to make a more traditional earthquake detection workflow applicable to LFEs.
The critical point for this is a viable automatic phase picker for LFE arrivals.
We borrow from the recent breakthroughs in seismic phase picking with deep learning, where recent neural network models have substantially improved earthquake detection \cite{zhu2019phasenet,ross2018generalized,munchmeyer2022picker}.
These neural network models are trained on millions of manually labelled phase arrivals and thereby learn to accurately discern seismic phase arrivals from noise and accurately determine arrival times.
The application of these models to continuous data has allowed to substantially increase the completeness of earthquake catalogs \cite{tan2021machine,gonzalez2023relation,moutote2023evidence}.

For tremor and LFE detection with deep learning, only few studies exist.
\citeA{rouet2020probing} identify tremor episodes in single-station records, but do not attempt to detect or locate individual LFEs.
\citeA{thomas2021identification} focus on LFEs on the San Andreas fault and test model configurations on cataloged events.
The preprint of \citeA{lin2023} presents an LFE detection workflow similar to the one presented here but focus exclusively on Southern Vancouver Island.
Here, we develop a deep learning based LFE picker and show its applicability to three independent study regions: Cascadia, Guerrero and Nankai.
To train our picker, we develop a novel strategy for synthetic data generation that allows for fine-grained control of the training process.
Using this method, we set up a classical earthquake detection workflow and demonstrate how to automatically create LFE catalogs across different world regions.
Our model successfully identifies and locates individual LFEs, even without using any training examples from the target region.
The resulting catalogs are coherent with classical catalogs but have been obtained in a fully automated and region-agnostic manner.
Furthermore, the catalogs identify LFEs missing from the reference catalogs, showing that our approach can uncover sources missed in the template matching procedures.
We make the trained picker available with a user-friendly interface through SeisBench \cite{woollam2022seisbench}.

\section{Training and validation of a deep learning LFE phase picker}

For detecting LFEs and determining their phase arrival times, we build a deep learning network.
Our network is closely modeled after PhaseNet \cite{zhu2019phasenet} due to the model's simplistic architecture and the excellent performance on earthquake data \cite{munchmeyer2022picker}.
PhaseNet is a 1D U-Net, i.e, a neural network consisting of convolutional encoder and decoder branches and skip connections \cite{ronneberger2015u}.
We provide the model with 60~s waveforms sampled at 20~Hz, bandpass-filtered between 1 and 8~Hz, the band in which LFEs are typically observed.
The model outputs probability curves for P and S phase arrivals.
We provide full details on the model and training procedure in the supplement.

\begin{figure*}[ht!]
\centering
\includegraphics[width=\textwidth]{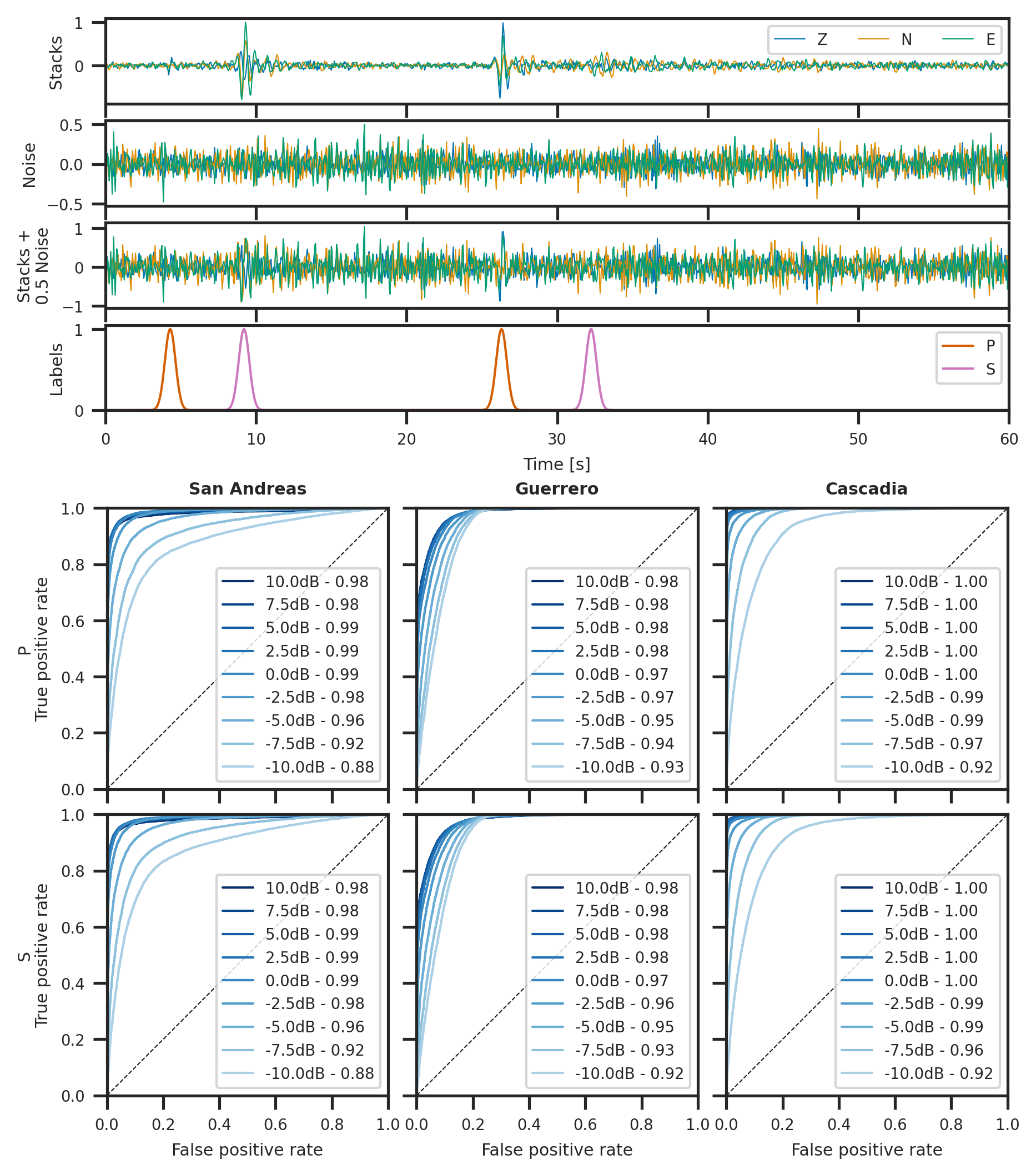}
\caption{Data generation procedure and evaluation results for synthetic data. The top panels show (top to bottom): the combination of two LFE stacks from Cascadia, a 60~s noise segment from INSTANCE, the combination of signal and noise at 3~dB SNR, and the Gaussian pulse labels for the P and S arrivals.
The bottom panels show the receiver operating characteristics (ROC) at different SNR. The numbers in the legend indicate the area under the ROC curve (AUC). For all plots, we use the joint model trained on all four datasets.}
\label{fig:synthetics_roc_combined}
\end{figure*}

In contrast to traditional earthquake pickers, training the model on cataloged LFE waveforms is suboptimal.
First, LFEs occur in bursts, i.e., around one LFE arrival there are often further arrivals many of which have not been labeled.
This leads to incorrect labelling and in addition makes a quantitative analysis of the model performance difficult.
Second, most LFE catalogs are based on template matching, i.e., individual arrivals need to be inferred from arrival times on templates.
Due to the low SNR, these times are often highly inaccurate, leading to high model uncertainties.
Instead, we train our model on synthetics.
For this, we combine LFE stacks with real seismic noise, allowing us to control the number and timing of LFEs and the SNR (Figure~\ref{fig:synthetics_roc_combined}).
We use up to three LFEs per trace to train the model to recognize events with low inter-event times.
We use seismic noise from the INSTANCE dataset for Italy \cite{michelini2021instance} as it contains no known LFEs.

We train our model using four regions: Southern Vancouver Island in Cascadia (Canada/USA) \cite{bostockMagnitudesMomentdurationScaling2015}, the central section of the San Andreas fault (USA) \cite{shelly15YearCatalog2017}, Guerrero (Mexico) \cite{frankUsingSystematicallyCharacterized2014}, and Nankai (Japan) \cite{jmabulletin}.
Figure~S1 shows the distribution of events and stations in the reference catalogs.
For Cascadia, San Andreas and Guerrero, we use template matching catalogs and the previously described strategy for generating examples.
For Nankai, we apply the classical training scheme as used for earthquakes as individual picks are available.
Further details on the datasets can be found in the supplement.

We evaluate our trained models quantitatively on synthetic examples generated with the previously described noise plus stack strategy.
The performance on synthetic data can serve as a proxy for the expected performance on real data.
We exclude the Nankai catalog from the analysis, because the catalog incompleteness precludes the extraction of challenging yet guaranteed LFE-free time windows.
As this study focuses on the generated LFE catalogs, we only provide a synopsis of the analysis on synthetics here and refer to the supplement for further details.

Overall, the models show excellent detection performance for both P and S waves, with an area under the curve (AUC) of the receiver operator characteristics of 0.97 to 1.00 in all regions for positive SNR in dB scale (Figure~\ref{fig:synthetics_roc_combined}).
The performance degrades mildly at -2.5~dB SNR and more sharply after, but all AUC values stay above 0.88 even at -10~dB SNR.
Models transfer well across regions with the worst results for a model trained exclusively on Cascadia (Figure~S2).
The best performing model is the one trained jointly on all four regions.
Therefore, we use the model trained jointly on all regions in the subsequent analysis unless explicitly stated otherwise.

Analysing the pick time residuals, clear regional differences are visible, with lowest residuals in Cascadia (Figure~S3).
In all regions, average residuals are about 0.2~s larger for P arrivals than S arrivals, indicating that these are more difficult to pick.
With standard deviations between 0.3 and 1.3~s (at 0~dB), residuals are substantially higher than for traditional earthquake pickers \cite{munchmeyer2022picker}.
Nonetheless, the residuals expose only low or no bias across all regions.
For the regional differences in performance, we think that they can primarily be attributed to the heterogeneity in data quality and SNR.
For example, the Cascadia stacks show the highest SNR, leading to the lowest pick residuals.
In turn, this implies that no conclusions about inherent regional differences in difficulty for picking LFEs can be inferred.

\section{Building deep learning LFE catalogs}

Using our phase picking model, we set up an LFE catalog workflow similar to the classical earthquake detection workflow.
Here we provide an overview of the workflow with further details in the supplement.
First, we pick P and S phases by applying the trained deep learning model to continuous waveforms using SeisBench \cite{woollam2022seisbench}.
Second, we use the PyOcto phase associator \cite{munchmeyer2023pyocto} to identify coherent arrivals across stations.
Third, we use NonLinLoc \cite{lomax2000probabilistic} with a 1D velocity model to perform absolute location of the events.
To avoid false detections, we filter the events based on the number of phase picks and the location residuals.
For comparison, we create earthquake catalogs using the same waveforms and workflow but using a PhaseNet model trained on INSTANCE implemented in SeisBench as the picker \cite{zhu2019phasenet,michelini2021instance,woollam2022seisbench}.
We remove events detected as both earthquakes and LFEs from the catalogs (Figure~S4).

\begin{figure*}[ht!]
\centering
\includegraphics[width=\textwidth]{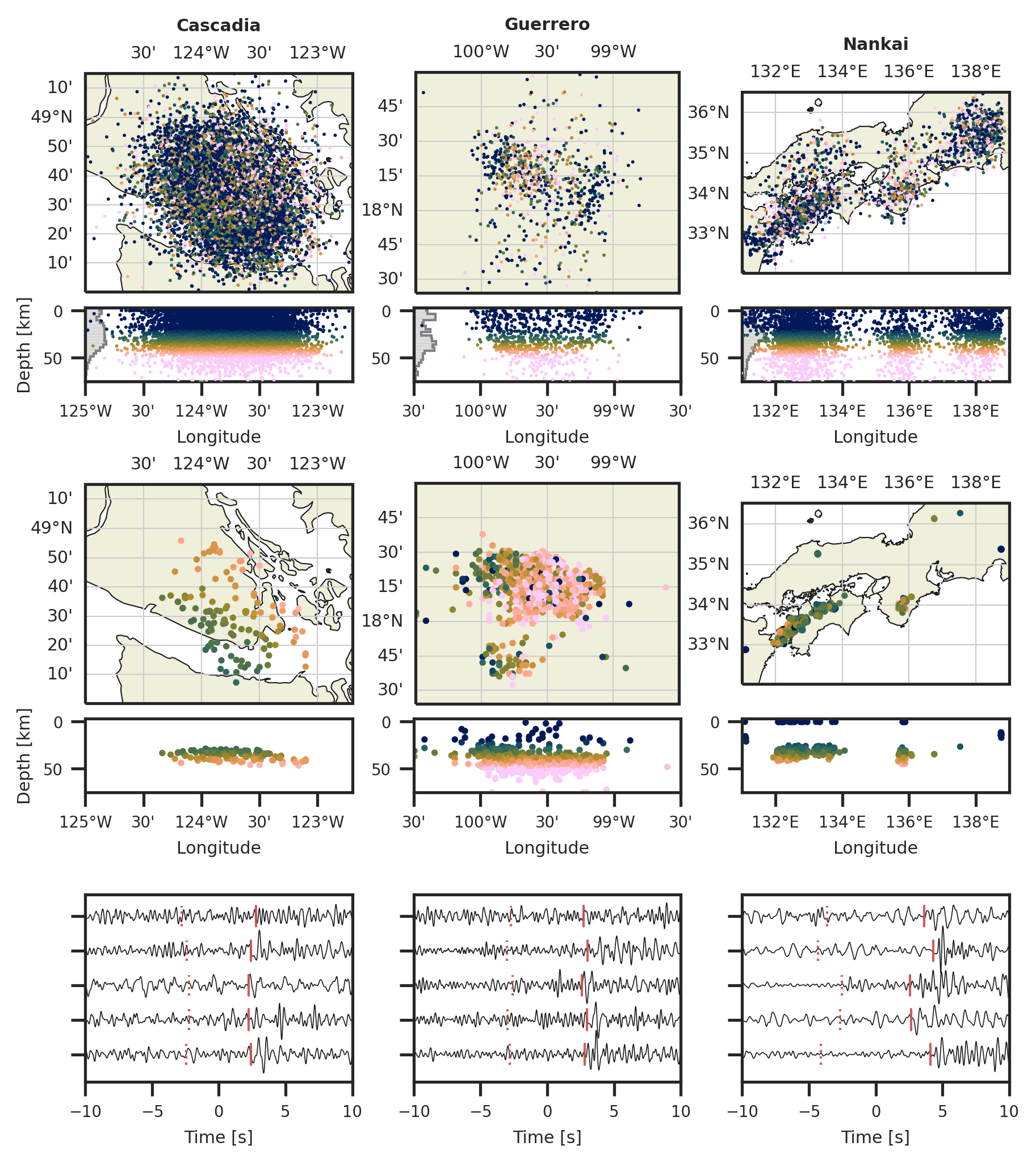}
\caption{LFE catalogs obtained using the deep learning (top row) and reference catalogs (bottom row). For deep learning, each dot represents one LFE. The bottom subpanels show depth cross-sections, showing longitude and depth of events. Color encodes event depth. The histograms on the left of the cross-section show the depth distribution of the detected events. In the reference catalogs for Cascadia \cite{bostockMagnitudesMomentdurationScaling2015} and Guerrero \cite{frankUsingSystematicallyCharacterized2014} each dot represents a LFE. For Nankai \cite{jmabulletin} individual LFEs are plotted. The bottom panel shows associated LFE picks from deep learning in each region. Red lines indicate phase picks (dotted for P, solid for S).}
\label{fig:map_joined}
\end{figure*}

We apply the workflow to compile LFE catalogs for the four study regions.
As we focus on studying the performance of the model and its resulting catalogs, we restrict ourselves to short study periods: 2003-02-26 to 2003-03-10, 2004-07-02 to 2004-07-27, and 2005-09-03 to 2005-09-25 for Cascadia; 2005-09-01 to 2005-11-30 for Guerrero; 2014-07-01 to 2014-10-01 for San Andreas; 2012-05-25 to 2012-06-14 for Nankai.
We chose the periods to contain both intense LFE activity and segments without any identified LFEs.

Figure~\ref{fig:map_joined} shows the spatial event distributions.
While the overall event locations are scattered, we notice strong similarities with the reference catalogs.
For Cascadia (8369 events detected), LFE activity is distributed almost evenly underneath South Vancouver island.
For Guerrero (772 events detected), LFEs occur mostly in a band between 100 and 99$^\circ$ West and around 18.25$^\circ$ North.
For Nankai (3120 events detected), a clear band of LFEs is visible in Southwestern Nankai.
Further LFEs around 135.5$^\circ$~E match the second band of LFEs commonly observed in Nankai.
For San Andreas (1074 events detected), the new catalog deviates from the previous observations, with the detection broadly distributed in space instead of along the fault (Figure~S5).
This is likely caused by very poor locations due to the station geometry.
As many events are only detected by the Parkfield borehole array with very dense station spacing, the aperture is small. 
Together with high pick uncertainties, this makes determining accurate locations challenging.
Therefore, we will exclude San Andreas from the following analysis.

In all catalogs, the event depth is poorly constrained.
While the largest density of events is around the previous estimates of LFE source depths, there is strong scatter.
We suggest that this is caused by the high timing uncertainty of the picks.
In particular, a high P pick uncertainty will cause poor depth constraints as the P to S time is indicative of depth.
Template matching catalogs alleviate this problem by locating LFE families instead of single events.
In Guerrero, we observe an arc-shaped depth distribution.
This is most likely related to the station distribution that traces out almost a straight line, leading to poor location constraints perpendicular to the station line \cite{frankAutomaticDetectionLowfrequency2014}.

\begin{figure*}[ht!]
\centering
\includegraphics[width=\textwidth]{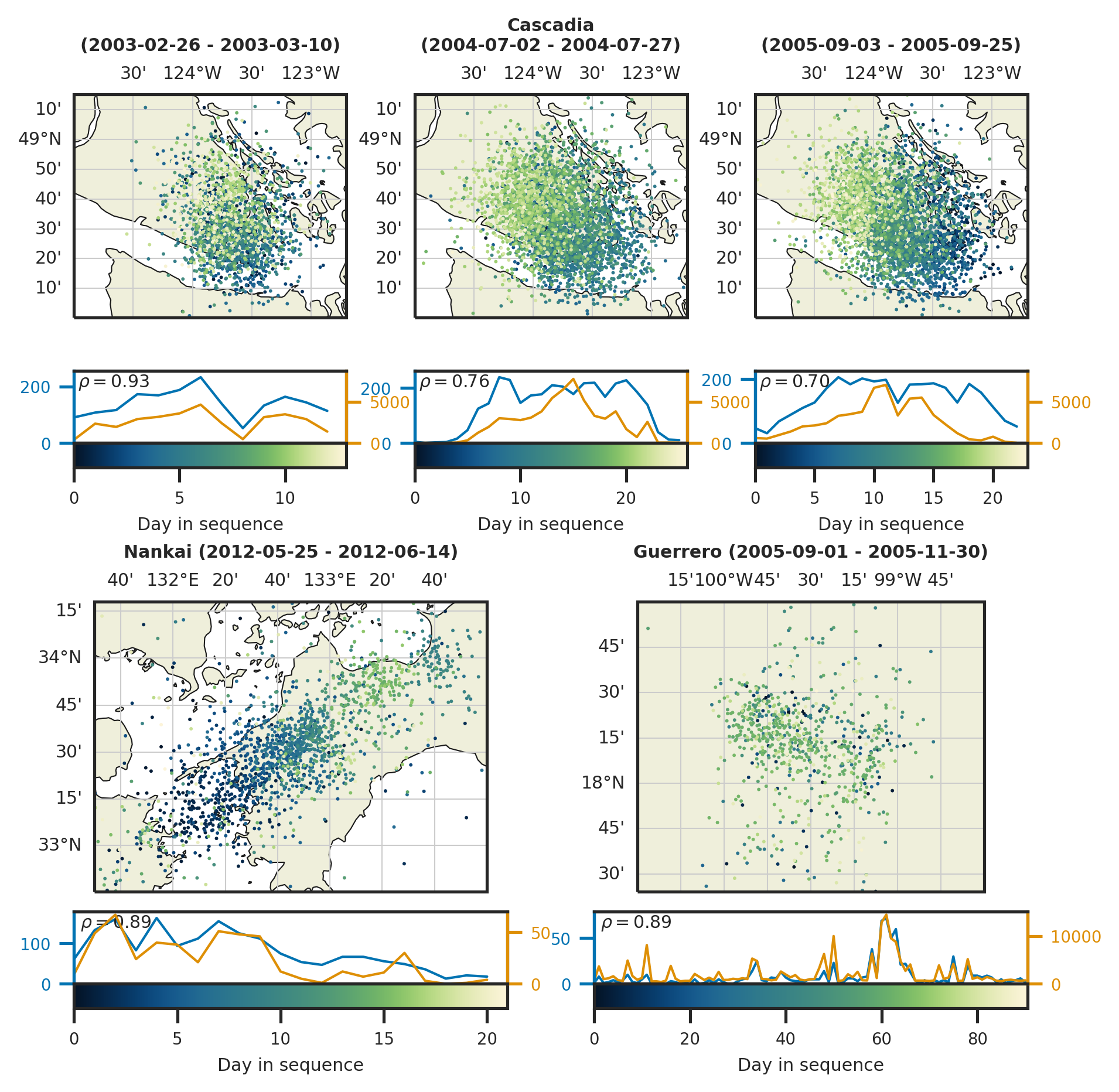}
\caption{Spatial and temporal migration patterns of detected LFEs. Each dot represents one LFE. The time within the sequence is indicated by color. The bottom panel shows the number of events per day for the LFEs from deep learning (blue) and reference catalogs (orange). The numbers in the upper left corners indicate the Pearson correlation coefficient between the daily number of events between the two catalogs.}
\label{fig:map_migrations}
\end{figure*}

Even though the overall shape of the catalogs is consistent with the previous catalogs, this alone does not confirm that the identified events are indeed LFEs.
We therefore conduct additional analysis into the newly obtained catalogs.
Figure~\ref{fig:map_migrations} shows spatial and temporal patterns in the catalogs.
In Cascadia, LFEs in all three observed sequences show a clear North-Westward migration.
This is consistent with typical slow slip and tremor migration patterns in the area \cite{wechCatalogingTectonicTremor2021}.
For Nankai, the observed migration is North-Eastward.
Notably, the migration is not continuous but rather has a gap and an additional, earlier cluster at the far North-West.
This pattern matches exactly the migration pattern in the JMA catalog.
We do not observe clear spatial migration patterns in Guerrero, however, such patterns have previously only been identified with very precise location estimates \cite{frankUsingSystematicallyCharacterized2014}.
In all regions, the evolution of daily event rate between the deep learning and reference catalogs is highly similar with Pearson correlation coefficients between 0.70 and 0.93.
In absolute numbers, the deep learning method detects substantially fewer events than the template matching, but more events than the manual detection procedure of the JMA.
We note that the number of events from deep learning is highly dependent on the chosen quality control parameters, which we set rather conservatively to avoid false positive detections.

\begin{figure*}[ht!]
\centering
\includegraphics[width=\textwidth]{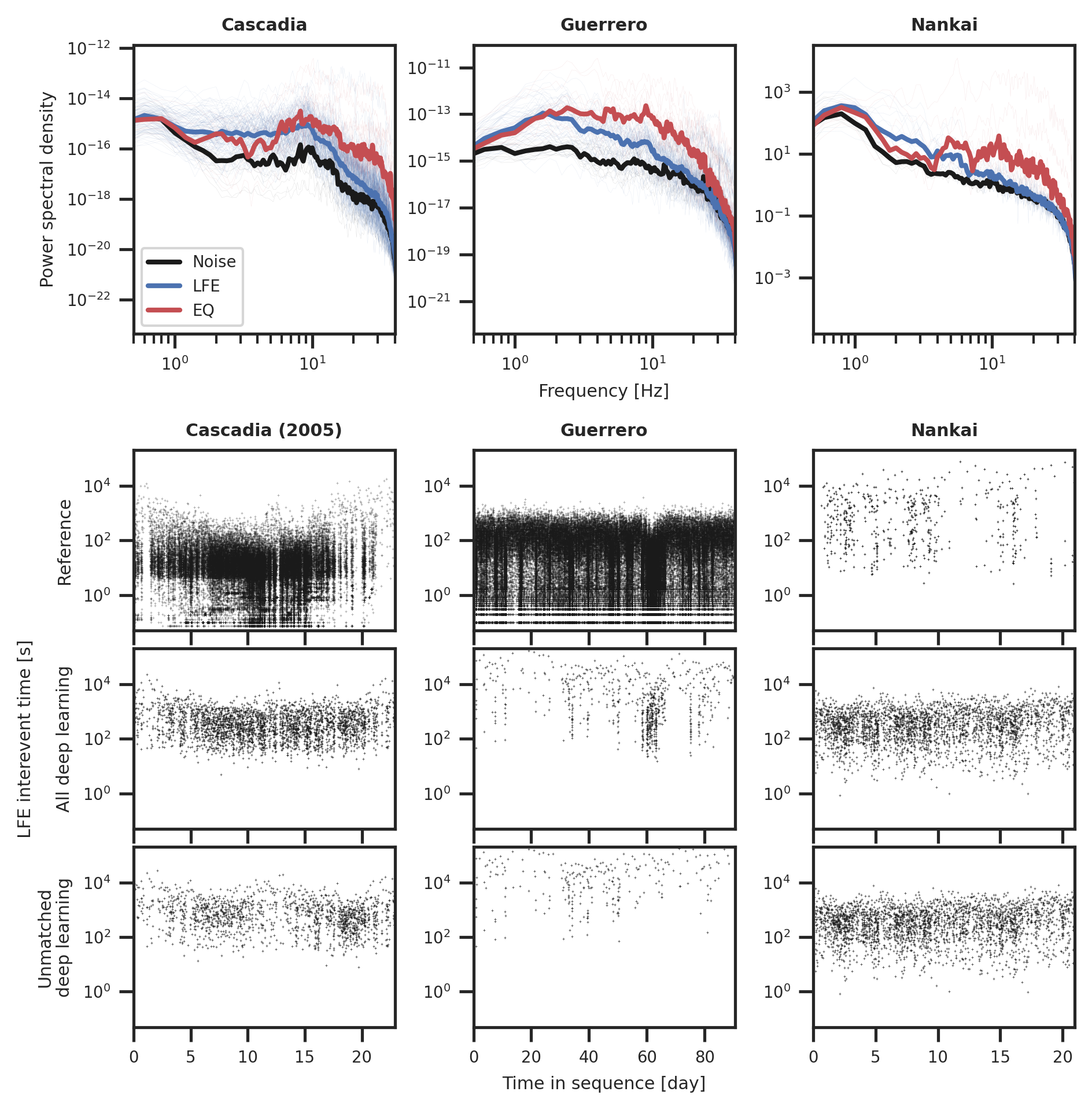}
\caption{Spectra and interevent times in the different regions. The top part shows velocity power spectral density for noise, earthquakes (EQs), and LFEs detected using deep learning. For each region, all traces stem from one reference station (Cascadia - MGCB, Guerrero - MAXE, Nankai - IIDH). Noise example have been extracted outside tremor episodes. Spectra have been calculated from the horizontal components (20~s windows for noise, 11~s windows starting 1~s before the S  arrival for events). Thin lines show individual spectra, bold lines median spectra. EQs were selected at a similar distance and depth range to the LFEs. Network averaged spectra are shown in Figure~S7.
The bottom part shows the development of interevent times during the LFE sequences in the reference catalogs, the deep learning catalogs, and for the unmatched events, i.e., all events from the deep learning catalogs that are not in the reference catalogs. Vertical stripes in the events indicate the occurrence of LFE bursts. For Cascadia, we only visualise the 2005 sequence for simplicity. We visualise all events from the reference catalogs without further declustering, leading to very low interevent times.}
\label{fig:spectra_interevent_combined}
\end{figure*}

Figure~\ref{fig:spectra_interevent_combined} shows a comparison of the velocity spectra of LFEs detected by our method, earthquakes and noise in the three regions.
The spectra clearly show the characteristics of the different event types.
While the earthquake spectra show increasing or at least constant energy up to about 10~Hz, the LFEs show a continuous decrease in energy from low frequencies onward.
The LFEs only show substantially higher energy than the noise in a small frequency band, while the EQs show substantially higher SNR at high frequencies.
This depletion in energy at higher frequencies is the key property of LFEs.

\section{Increased diversity of LFE sources through deep learning}

We compare the detected events to the reference catalogs (Figure~S6).
In Cascadia, for 60~\% of the LFEs from our workflow, we find an LFE in the reference catalog within 10~s; for Guerrero for 70~\% of the events.
For Nankai, only 6\% of our LFEs are in the reference catalog, however our catalog also substantially exceeds the original catalog in the total number of events.
Conversely, we recover 32\% of the cataloged events.
Note that a loose threshold for matching is justified due to uncertainties in the origin times due to inaccurate locations.
On one hand, these results are another confirmation that the method correctly identifies LFEs.
On the other hand, the substantial fraction of uncataloged events suggest that our method identifies previously unidentified LFEs.
In the following, we verify and analyse these detections.

First, we rule out spurious detection.
To this end, we scramble the picked arrival times of each station by applying small random shifts.
We choose constant shifts for each station for every hour.
This destroys the exact times, while keeping the pick distribution, the P to S times within a station, and the higher number of picks within tremor bursts intact.
The scrambled ``catalogs'' only contain about 5 to 10~\% of the number of events contained in the original catalogs and show no spatial coherence (Figure~S8).
Even these numbers are still likely an overestimation of the false positive rate, as events recorded at many stations are likely to be unperturbed by our scrambling procedure.
The more than 30~\% additional detections can therefore not be attributed to spurious associations.

Mapping the events without matches in the reference catalog (Figure~S9) reveals that they follow the same spatial extent and migration pattern as the full catalog.
Notably, for Cascadia and Mexico there are changes in the temporal patterns.
For Cascadia, the newly detected events concentrate early and late in the LFE sequence, coinciding with a spatial location around the southeastern tip of Vancouver island and towards the northwestern end of the LFE cluster.
For Mexico, the largest fraction of new detections clusters in time between days 30 and 50 of the analyses sequence.
Visualising the interevent time (Figure \ref{fig:spectra_interevent_combined}) confirms these observations.
Both the full deep learning catalogs and the catalog of events without a match in the reference catalogs show clear burst behaviour.
In particular for Mexico, certain LFE bursts are contained virtually completely in the template matching catalog, while others have not been identified at all.
This highlights that the newly detected LFEs do not only uncover new sources but even new LFE bursts.

To further validate this finding, we correlate the uncataloged detections with the family stacks from the reference catalogs (Figure~S10).
For Cascadia, the distribution of correlation values for these uncataloged detections are identical to the noise distribution, i.e., the new detections do not match known sources.
For Guererro, some events show systematically higher correlations than the noise.
Nonetheless, the vast majority of our additional detections does not match the known sources any better than the noise.
This verifies that these new detections are systematically different and that these events can not be found with template matching without identifying further, novel templates.

Extending upon the finding that the model can generalise from known families to LFEs outside these families, we investigate the ability to detect LFEs in regions the model has not been trained on.
For this analysis, we trained leave-one-out models, i.e., models trained on all but one region, and applied them on the left-out region.
Figure~S11 visualises the spatial and temporal migration patterns.
Again, the clear migration patterns in Nankai and Cascadia are retrieved.
Furthermore, the number of events correlates highly (Pearson correlation between 0.70 and 0.90) with the reference catalogs.
The total size of the catalogs varies, with a substantially smaller catalog in Cascadia, a similarly-sized catalog in Nankai, and a far bigger catalog in Guerrero.
However, these might be related to changes in the model confidence values rather than their actual quality as we produced all catalogs with fixed picking thresholds.
The cross-regional analysis clearly illustrates that the models can be transferred across regions and recover LFEs from families they have not been trained on.

\section{Conclusion}

Our analysis shows that deep learning and template matching are complementary in the way they detect LFEs with specific advantages and disadvantages for either method.
The biggest strength of deep learning is the flexibility.
The model can be applied to additional stations, including temporary stations, shows higher diversity in terms of event families, and can be transferred across regions.
In addition, it is able to detect and locate single LFEs, even though with substantial uncertainties in terms of depth.
In contrast, template matching requires a predefined set of sources that is difficult to obtain and specific to each region and set of stations.
While rigid, this leads to a more sensitive model, as evidenced by higher event counts.
Furthermore, it allows template matching to identify LFEs with fewer stations than deep learning.
A promising avenue might be the combination of deep learning and template matching, i.e., using deep learning to identify a diverse set of templates and then use template matching to increase the completeness of the identified families.

Lastly, the deep learning method extends our view of LFEs by detecting previously unidentified sources.
Building a comprehensive set of templates for template matching is challenging: the low bandwidth and SNR makes it difficult to distinguish between closely spaces sources, leading to a trade-off between missing sources and redundant templates.
In contrast, the deep learning method is source-agnostic, i.e., no selection of sources needs to be performed for detecting individual events.
Such a source-agnostic view is necessary to perform unbiased subsequent analysis that require a complete view of LFE sources, such as estimates of slow slip.
In addition, the fact that the model can be transferred across regions shows that LFEs have universal, region-independent properties similar to earthquakes.
Given our results, we expect that deep learning methods will allow to map LFEs across world regions with high consistency and diversity.

\section*{Open Research Section}
The stack data sets for Cascadia, San Andreas, and Guerrero are available through SeisBench (\url{https://github.com/seisbench/seisbench}, \url{https://doi.org/10.5281/zenodo.5568812}) from version v0.x onwards.
The model including pretrained weights is available through SeisBench starting from the same version.\footnote{Data and models will be made available upon acceptance.}
The INSTANCE noise dataset is available through SeisBench and at \url{https://doi.org/10.13127/instance}.
The FNet data is available from the NIED at \url{https://www.fnet.bosai.go.jp/}.
We use waveforms from the BK \cite{fdsnBK}, BP \cite{fdsnBP}, CI \cite{fdsnCI}, CN \cite{fdsnCN}, C8 \cite{fdsnC8}, NC \cite{fdsnNC}, PB \cite{fdsnPB}, PO \cite{fdsnPO}, TA \cite{fdsnTA}, and TO (MASE deployment) \cite{fdsnTO} networks.
Waveforms were accessed through the IRIS and NCEDC FDSN webservices.

\acknowledgments
We thank David Shelly for sharing his data on the San Andreas LFE catalog.
We thank Albanne Lecointre for help with processing the FNet data.
This work has been supported by MIAI@Grenoble Alpes (ANR-19-P3IA-0003).
This work has been partially funded by the European Union under the grant agreement n°101104996 (“DECODE”) and the ERC CoG 865963 DEEP-trigger. Views and opinions expressed are however those of the authors only and do not necessarily reflect those of the European Union or REA. Neither the European Union nor the granting authority can be held responsible for them.
This work was granted access to the HPC/AI resources of IDRIS under the allocation 2022-AD011012345R1 made by GENCI.

\bibliography{zotero,mybibfile}

\end{document}